\documentclass[aps,prd,showpacs,preprint]{revtex4}
\usepackage{amssymb}
\usepackage{amsmath}
\usepackage{graphicx}

\begin{document}

\title{Jacobi stability of the vacuum in the static spherically symmetric brane world models}

\author{T.~Harko}
\email{harko@hkucc.hku.hk} \affiliation{Department of Physics and
Center for Theoretical
             and Computational Physics, The University of Hong Kong,
             Pok Fu Lam Road, Hong Kong}

\author{V.~S.~Sabau}
\email{sorin@cc.htokai.ac.jp} \affiliation{Department of
Mathematics, Hokkaido Tokai University, 5-1-1-1 Minamisawa,
Minami-ku, Sapporo, 005-8601, Japan}

\date{\today}

\begin{abstract}

We analyze the stability of the structure equations of the vacuum
in the brane world models, by using both the linear (Lyapunov)
stability analysis, and the Jacobi stability analysis, the
Kosambi-Cartan-Chern (KCC) theory. In the brane world models the
four dimensional effective Einstein equations acquire extra terms,
called dark radiation and dark pressure, respectively, which arise
from the embedding of the 3-brane in the bulk. Generally, the
spherically symmetric vacuum solutions of the brane gravitational
field equations, have properties quite distinct as compared to the
standard black hole solutions of general relativity. We close the
structure equations by assuming a simple linear equation of state
for the dark pressure. In this case the vacuum is Jacobi stable
only for a small range of values of the proportionality constant
relating the dark pressure and the dark radiation. The unstable
trajectories on the brane behave chaotically, in the sense that
after a finite radial distance it would be impossible to
distinguish the trajectories that were very near each other at an
initial point. Hence the Jacobi stability analysis offers a
powerful method for constraining the physical properties of the
vacuum on the brane.
\end{abstract}

\pacs{04.50.-h, 11.10.Kk, 02.30.Oz, 02.30.Hq}

\maketitle

\section{Introduction}

The idea of embedding our Universe in a higher dimensional space has
attracted a considerable interest recently, due to the proposal by Randall
and Sundrum that our four-dimensional (4D) spacetime is a three-brane,
embedded in a 5D spacetime (the bulk) ~\cite{RS99a}. According to the brane
world scenario, the physical fields (electromagnetic, Yang-Mills etc.) in
our 4D Universe are confined to the three brane. Only gravity can freely
propagate in both the brane and bulk spacetimes, with the gravitational
self-couplings not significantly modified. Even if the fifth dimension is
uncompactified, standard 4D gravity is reproduced on the brane. Hence this
model allows the presence of large, or even infinite non-compact extra
dimensions. Our brane is identified to a domain wall in a 5D anti-de Sitter
spacetime. In the brane world scenario, the fundamental scale of gravity is
not the Planck scale, but another scale which may be at the TeV level (for a
review of the dynamics and geometry of brane world models see~\cite{Mart04}%
). Due to the correction terms coming from the extra dimensions, significant
deviations from the standard Einstein theory occur at very high energies~
\cite{SMS00}. Gravity is largely modified at the electro-weak scale of about
$1$~TeV. The cosmological and astrophysical implications of the brane world
theories have been extensively investigated in the physical literature~\cite
{all2,sol}.

The static vacuum gravitational field equations on the brane depend on the
generally unknown Weyl stresses, which can be expressed in terms of two
functions, called the dark radiation $U$ and the dark pressure $P$ terms
(the projections of the Weyl curvature of the bulk, generating non-local
brane stresses)~\cite{Da00,Mart04,GeMa01}.

Several classes of spherically symmetric solutions of the static
gravitational field equations in the vacuum on the brane have been obtained
in~\cite{Ha03,Ma04,Ha05}. As a possible physical application of these
solutions the behavior of the angular velocity $v_{tg}$ of the test
particles in stable circular orbits has been considered~\cite
{Ma04,Ha05,BoHa07,HaCh07}. The observed properties of the rotational
galactic curves ~\cite{Bi87} can be naturally explained in the brane world
models, without introducing any additional hypothesis. The galaxy is
embedded in a modified, spherically symmetric geometry, generated by the
non-zero contribution of the Weyl tensor from the bulk. The extra terms,
which can be described in terms of the dark radiation term $U$ and the dark
pressure term $P$, act as a ``matter'' distribution outside the galaxy. The
particles moving in this geometry feel the gravitational effects of $U$,
which can be expressed in terms of an equivalent mass $M_{U}$ (the dark
mass)~\cite{Ma04,BoHa07,HaCh07}. The role of the cross-over length scale in
the possible explanation of the dark-matter phenomenon in the brane world
model was investigated in \cite{VySh07}. Similar interpretations of the dark
matter as a bulk effect have been also considered in~\cite{Pal}.

For standard general relativistic spherical compact objects the exterior
space-time is described by the Schwarzschild metric. In the five dimensional
brane world models, the high energy corrections to the energy density,
together with the Weyl stresses from bulk gravitons, imply that on the brane
the exterior metric of a static star is no longer the Schwarzschild metric
\cite{Da00}. The presence of the Weyl stresses also means that the matching
conditions do not have a unique solution on the brane; the knowledge of the
five-dimensional Weyl tensor is needed as a minimum condition for uniqueness.

Static, spherically symmetric exterior vacuum solutions of the brane world
models have been obtained first in \cite{Da00} and in \cite{GeMa01}. The
first of these solutions has the mathematical form of the Reissner-Nordstrom
solution of the standard general relativity, in which a tidal Weyl parameter
plays the role of the electric charge of the general relativistic solution
\cite{Da00}. A second exterior solution, which also matches a constant
density interior, has been derived in \cite{GeMa01}. Other vacuum solutions
of the field equations in the brane world model have been obtained in \cite
{cfm02,Da03,Vi03,Ca03,BMD03}.

Generally, the vacuum field equations on the brane can be reduced to a
system of two ordinary differential equations, which describe all the
geometric properties of the vacuum as functions of the dark pressure and
dark radiation terms~\cite{Ha03}. In order to close the system of vacuum
field equations on the brane a functional relation between these two
quantities is necessary.

Hence, a first possible approach to the study of the vacuum brane consists
in adopting an explicit equation of state for the dark pressure as a
function of the dark radiation. The explicit form of this equation can be
obtained by assuming that the brane obeys different geometrical or physical
conditions. The existence of the group of the homology transformations on
the brane uniquely fixes the form of the equation of state \cite{Ha03}). By
assuming that the vacuum admits the group of conformal motions leads to the
full determination of the equation of state of the dark pressure \cite
{Ha03,Ma04}. By imposing the condition of the constancy of the rotational
velocity curves for particle in stable orbits closes the system of the field
equations, and the dark radiation and pressure can be obtained explicitly
\cite{Ha03,Ha05,BoHa07,HaCh07}. This approach has the advantage of making
predictions that may be tested directly by observations.

It is the purpose of the present paper to consider an alternative
possibility for constraining the equation of state of the dark pressure on
the brane. The structure equations of the vacuum can be transformed to an
autonomous system of two differential equations, which in turn may be
reduced to a single, second order differential equation. A second order
differential equation can be investigated in geometric terms by using the
general path-space theory of Kosambi-Cartan-Chern (KCC-theory)
inspired by the geometry of a 
Finsler space \cite{Ko33,Ca33,Ch39}. The KCC theory is a differential
geometric theory of the variational equations for the deviation of the whole
trajectory to nearby ones. By associating a non-linear connection and a
Berwald type connection to the differential system, five geometrical
invariants are obtained. The second invariant gives the Jacobi stability of
the system \cite{Sa05,Sa05a}. The KCC theory has been applied for the study
of different physical, biochemical or technical systems (see \cite{Sa05,
Sa05a, An93, An00} and references therein).

As a toy model for the applications of the KCC theory for the study of the
Jacobi stability of the vacuum in the brane world models we consider the
case of the linear equation of state for the dark pressure, $P=\gamma U$.
The vacuum on the brane is Jacobi stable only for a very limited range of
the proportionality constant $\gamma $, $-0.5\leq \gamma \leq 0.67$, and it
is Jacobi unstable for all the other possible values of $\gamma $. Hence the
natural physical requirement of Jacobi stability provides a very strong
constraint on the equation of state of the dark pressure, and on the
physical/geometrical properties of the model. Such a constraint cannot be
obtained from the linear (Lyapunov) stability analysis, which is also
considered in detail.

The present paper is organized as follows. The field equations for the
vacuum on the brane are written down in Section II. The structure equations
of the vacuum are derived in Section III. We review the mathematical
formalism of the KCC theory in Section IV. The linear stability analysis of
the structure equations of the vacuum for a linear equation of state for the
dark pressure is performed in Section V. The Jacobi stability of the
structure equations is analyzed in Section VI. We discuss and conclude our
results in Section VII.

\section{The field equations for static, spherically symmetric vacuum branes}

\label{field}

In the present Section we briefly describe the basic mathematical formalism
of the brane world models, present the field equations for a static,
spherically symmetric vacuum brane, and discuss some of their consequences.

\subsection{The field equations in the brane world models}

We start by considering a five dimensional (5D) spacetime (the bulk), with a
single four-dimensional (4D) brane, on which matter is confined. The 4D
brane world $({}^{(4)}M,g_{\mu \nu })$ is located at a hypersurface $\left(
B\left( X^{A}\right) =0\right) $ in the 5D bulk spacetime $%
({}^{(5)}M,g_{AB}) $, of which coordinates are described by $%
X^{A},A=0,1,...,4$. The induced 4D coordinates on the brane are $x^{\mu
},\mu =0,1,2,3$.

The action of the system is given by~\cite{SMS00}
\begin{equation}
S=S_{bulk}+S_{brane},  \label{bulk}
\end{equation}
where
\begin{equation}
S_{bulk}=\int_{{}^{(5)}M}\sqrt{-{}^{(5)}g}\left[ \frac{1}{2k_{5}^{2}}{}%
^{(5)}R+{}^{(5)}L_{m}+\Lambda _{5}\right] d^{5}X,
\end{equation}
and
\begin{equation}
S_{brane}=\int_{{}^{(4)}M}\sqrt{-{}^{(5)}g}\left[ \frac{1}{k_{5}^{2}}K^{\pm
}+L_{brane}\left( g_{\alpha \beta },\psi \right) +\lambda _{b}\right] d^{4}x,
\end{equation}
where $k_{5}^{2}=8\pi G_{5}$ is the 5D gravitational constant, ${}^{(5)}R$
and ${}^{(5)}L_{m}$ are the 5D scalar curvature and the matter Lagrangian in
the bulk, $L_{brane}\left( g_{\alpha \beta },\psi \right) $ is the 4D
Lagrangian, which is given by a generic functional of the brane metric $%
g_{\alpha \beta }$ and of the matter fields $\psi $, $K^{\pm }$ is the trace
of the extrinsic curvature on either side of the brane, and $\Lambda _{5}$
and $\lambda _{b}$ (the constant brane tension) are the negative vacuum
energy densities in the bulk and on the brane, respectively.

The Einstein field equations in the bulk are given by~\cite{SMS00}
\begin{equation}
{}^{(5)}G_{IJ}=k_{5}^{2} {}^{(5)}T_{IJ},\qquad {}^{(5)}T_{IJ}=-\Lambda _{5}
{}^{(5)}g_{IJ}+\delta(B)\left[-\lambda_{b} {}^{(5)}g_{IJ}+T_{IJ}\right] ,
\end{equation}
where
\begin{equation}
{}^{(5)}T_{IJ}\equiv - 2\frac{\delta {}^{(5)}L_{m}}{\delta {}^{(5)}g^{IJ}}
+{}^{(5)}g_{IJ} {}^{(5)}L_{m},
\end{equation}
is the energy-momentum tensor of bulk matter fields, while $T_{\mu \nu }$ is
the energy-momentum tensor localized on the brane and which is defined by
\begin{equation}
T_{\mu \nu }\equiv -2\frac{\delta L_{brane}}{\delta g^{\mu \nu }}+g_{\mu \nu
}\text{ }L_{brane}.
\end{equation}

The delta function $\delta \left( B\right) $ denotes the localization of
brane contribution. In the 5D spacetime a brane is a fixed point of the $%
Z_{2}$ symmetry. The basic equations on the brane are obtained by
projections onto the brane world. The induced 4D metric is $%
g_{IJ}={}^{(5)}g_{IJ}-n_{I}n_{J}$, where $n_{I}$ is the space-like unit
vector field normal to the brane hypersurface ${}^{(4)}M$. In the following
we assume ${}^{(5)}L_{m}=0$. In the brane world models only gravity can
probe the extra dimensions.

Assuming a metric of the form $ds^{2}=(n_{I}n_{J}+g_{IJ})dx^{I}dx^{J}$, with
$n_{I}dx^{I}=d\chi $ the unit normal to the $\chi =\mathrm{constant}$
hypersurfaces and $g_{IJ}$ the induced metric on $\chi =\mathrm{constant}$
hypersurfaces, the effective 4D gravitational equation on the brane takes
the form~\cite{SMS00}:
\begin{equation}
G_{\mu \nu }=-\Lambda g_{\mu \nu }+k_{4}^{2}T_{\mu \nu }+k_{5}^{4}S_{\mu \nu
}-E_{\mu \nu },  \label{Ein}
\end{equation}
where $S_{\mu \nu }$ is the local quadratic energy-momentum correction
\begin{equation}
S_{\mu \nu }=\frac{1}{12}TT_{\mu \nu }-\frac{1}{4}T_{\mu }{}^{\alpha }T_{\nu
\alpha }+\frac{1}{24}g_{\mu \nu }\left( 3T^{\alpha \beta }T_{\alpha \beta
}-T^{2}\right) ,
\end{equation}
and $E_{\mu \nu }$ is the non-local effect from the free bulk gravitational
field, the transmitted projection of the bulk Weyl tensor $C_{IAJB}$, $%
E_{IJ}=C_{IAJB}n^{A}n^{B}$, with the property $E_{IJ}\rightarrow E_{\mu \nu
}\delta _{I}^{\mu }\delta _{J}^{\nu }\quad $as$\quad \chi \rightarrow 0$. We
have also denoted $k_{4}^{2}=8\pi G$, with $G$ the usual 4D gravitational
constant.

The 4D cosmological constant, $\Lambda $, and the 4D coupling constant, $%
k_{4}$, are related by $\Lambda =k_{5}^{2}(\Lambda _{5}+k_{5}^{2}\lambda
_{b}^{2}/6)/2$ and $k_{4}^{2}=k_{5}^{4}\lambda _{b}/6$, respectively. In the
limit $\lambda _{b}^{-1}\rightarrow 0$ we recover standard general
relativity \cite{SMS00}.

The Einstein equation in the bulk and the Codazzi equation also imply the
conservation of the energy-momentum tensor of the matter on the brane, $%
D_{\nu }T_{\mu }{}^{\nu }=0$, where $D_{\nu }$ denotes the brane covariant
derivative. Moreover, from the contracted Bianchi identities on the brane it
follows that the projected Weyl tensor obeys the constraint $D_{\nu }E_{\mu
}{}^{\nu }=k_{5}^{4}D_{\nu }S_{\mu }{}^{\nu }$.

The symmetry properties of $E_{\mu \nu }$ imply that in general we can
decompose it irreducibly with respect to a chosen $4$-velocity field $%
u^{\mu} $ as~\cite{Mart04}
\begin{equation}
E_{\mu \nu }=-k^{4}\left[ U\left( u_{\mu }u_{\nu} +\frac{1}{3}h_{\mu \nu
}\right) +P_{\mu \nu }+2Q_{(\mu }u_{\nu)}\right] ,  \label{WT}
\end{equation}
where $k=k_{5}/k_{4}$, $h_{\mu \nu }=g_{\mu \nu }+u_{\mu }u_{\nu }$ projects
orthogonal to $u^{\mu }$, the ``dark radiation'' term $U=-k^{-4}E_{\mu \nu
}u^{\mu }u^{\nu }$ is a scalar, $Q_{\mu }=k^{-4}h_{\mu }^{\alpha }E_{\alpha
\beta }u^{\beta }$ is a spatial vector and $P_{\mu \nu }=-k^{-4}\left[
h_{(\mu }\text{ }^{\alpha }h_{\nu )}\text{ }^{\beta }-\frac{1}{3}h_{\mu \nu
}h^{\alpha \beta }\right] E_{\alpha \beta }$ is a spatial, symmetric and
trace-free tensor.

In the case of the vacuum state we have $\rho =p=0$, $T_{\mu \nu }\equiv 0$,
and consequently $S_{\mu \nu }\equiv 0$. Therefore the field equation
describing a static brane takes the form
\begin{equation}
R_{\mu \nu }=-E_{\mu \nu }+\Lambda g_{\mu \nu },
\end{equation}
with the trace $R$ of the Ricci tensor $R_{\mu \nu }$ satisfying the
condition $R=R_{\mu }^{\mu }=4\Lambda $.

In the vacuum case $E_{\mu \nu }$ satisfies the constraint $D_{\nu }E_{\mu
}{}^{\nu }=0$. In an inertial frame at any point on the brane we have $%
u^{\mu }=\delta _{0}^{\mu }$ and $h_{\mu \nu }=\mathrm{diag}(0,1,1,1)$. In a
static vacuum $Q_{\mu }=0$ and the constraint for $E_{\mu \nu }$ takes the
form~ \cite{GeMa01}
\begin{equation}
\frac{1}{3}D_{\mu }U+\frac{4}{3}UA_{\mu }+D^{\nu }P_{\mu \nu }+A^{\nu
}P_{\mu \nu }=0,
\end{equation}
where $A_{\mu }=u^{\nu }D_{\nu }u_{\mu }$ is the 4-acceleration. In the
static spherically symmetric case we may chose $A_{\mu }=A(r)r_{\mu }$ and $%
P_{\mu \nu }=P(r)\left( r_{\mu }r_{\nu }-\frac{1}{3}h_{\mu \nu }\right) $,
where $A(r)$ and $P(r)$ (the ``dark pressure'') are some scalar functions of
the radial distance~$r$, and~$r_{\mu }$ is a unit radial vector~\cite{Da00}.

\subsection{The gravitational field equations for a static spherically
symmetric brane}

In the following we will restrict our study to the static and
spherically symmetric metric given by
\begin{equation}
ds^{2}=-e^{\nu (r)}dt^{2}+e^{\lambda (r)}dr^{2}+r^{2}\left(d\theta ^{2}+\sin
^{2}\theta d\phi ^{2}\right).  \label{metr1}
\end{equation}

With the metric given by~(\ref{metr1}) the gravitational field equations and
the effective energy-momentum tensor conservation equation in the vacuum
take the form~\cite{Ha03,Ma04}
\begin{equation}
-e^{-\lambda }\left( \frac{1}{r^{2}}-\frac{\lambda ^{\prime }}{r}\right) +%
\frac{1}{r^{2}}=3\alpha U+\Lambda ,  \label{f1}
\end{equation}
\begin{equation}
e^{-\lambda }\left( \frac{\nu ^{\prime }}{r}+\frac{1}{r^{2}}\right) -\frac{1%
}{r^{2}}=\alpha \left( U+2P\right) -\Lambda ,  \label{f2}
\end{equation}
\begin{equation}
\frac{1}{2}e^{-\lambda }\left( \nu ^{\prime \prime }+\frac{\nu ^{\prime 2}}{2%
}+\frac{\nu ^{\prime }-\lambda ^{\prime }}{r}-\frac{\nu ^{\prime }\lambda
^{\prime }}{2}\right) =\alpha \left( U-P\right) -\Lambda ,  \label{f3}
\end{equation}
\begin{equation}
\nu ^{\prime }=-\frac{U^{\prime }+2P^{\prime }}{2U+P}-\frac{6P}{r\left(
2U+P\right) },  \label{f4}
\end{equation}
where $^{\prime }=d/dr$, and we have denoted $\alpha =16\pi G/k^{4}\lambda
_{b}$.

The field equations~(\ref{f1})--(\ref{f4}) can be interpreted as
describing
an isotropic ''matter distribution'', with the effective energy density $%
\rho ^{\mathrm{eff}}$, radial pressure $P^{\mathrm{eff}}$ and orthogonal
pressure $P_{\perp }^{\mathrm{eff}}$, respectively, so that $\rho ^{\mathrm{%
eff}}=3\alpha U+\Lambda $, $P^{\mathrm{eff}}=\alpha U+2\alpha P-\Lambda $
and $P_{\perp }^{\mathrm{eff}}=\alpha U-\alpha P-\Lambda $, respectively,
which gives the condition $\rho ^{\mathrm{eff}}-P^{\mathrm{eff}}-2P_{\perp
}^{\mathrm{eff}}=4\Lambda =\mathrm{constant}$. This is expected for the
`radiation' like source, for which the projection of the bulk Weyl tensor is
trace-less, $E_{\mu }^{\mu }=0$.

\section{Structure equations of the vacuum in the brane world models}

\label{Sect3}

\label{sec:3}Eq.~(\ref{f1}) can immediately be integrated to give
\begin{equation}
e^{-\lambda }=1-\frac{C_{1}}{r}-\frac{GM_{U}\left( r\right) }{r}-\frac{%
\Lambda }{3}r^{2},  \label{m1}
\end{equation}
where $C_{1}$ is an arbitrary constant of integration, and we denoted
\begin{equation}
GM_{U}\left( r\right) =3\alpha \int_{0}^{r}U(r)r^{2}dr.
\end{equation}

The function $M_U$ is the gravitational mass corresponding to the dark
radiation term (the dark mass). For $U=0$ the metric coefficient given by
Eq.~(\ref{m1}) must tend to the standard general relativistic Schwarzschild
metric coefficient, which gives $C_{1}=2GM$, where $M = \mathrm{constant}$
is the baryonic (usual) mass of the gravitating system.

By substituting $\nu ^{\prime }$ given by Eq.~(\ref{f4}) into Eq.~(\ref{f2})
and with the use of Eq.~(\ref{m1}) we obtain the following system of
differential equations satisfied by the dark radiation term $U$, the dark
pressure $P$ and the dark mass $M_{U}$, describing the vacuum gravitational
field, exterior to a massive body, in the brane world model \cite{Ha03}:
\begin{equation}
\frac{dM_{U}}{dr}=\frac{3\alpha }{G}r^{2}U.  \label{e2}
\end{equation}
\begin{equation}
\frac{dU}{dr}=-\frac{\left( 2U+P\right) \left[ 2GM+GM_{U}-\frac{2}{3}\Lambda
r^{3}+\alpha \left( U+2P\right) r^{3}\right] }{r^{2}\left( 1-\frac{2GM}{r}-%
\frac{M_{U}}{r}-\frac{\Lambda }{3}r^{2}\right) }-2\frac{dP}{dr}-\frac{6P}{r},
\label{e1}
\end{equation}

To close the system a supplementary functional relation between one of the
unknowns $U$, $P$ and $M_{U}$ is needed. Generally, this equation of state
is given in the form $P=P(U)$. Once this relation is known, Eqs.~(\ref{e2}%
)--(\ref{e1}) give a full description of the geometrical properties of the
vacuum on the brane.

In the following we will restrict our analysis to the case $\Lambda =0$.
Then the system of equations~(\ref{e2}) and~(\ref{e1}) can be transformed to
an autonomous system of differential equations by means of the
transformations
\begin{equation}
q=\frac{2GM}{r}+\frac{GM_{U}}{r},\qquad \mu =3\alpha r^{2}U,  \label{transf1}
\end{equation}
\begin{equation}
p=3\alpha r^{2}P,\qquad \theta =\ln r,  \label{transf2}
\end{equation}
where $\mu $ and $p$ are the ``reduced'' dark radiation and pressure,
respectively.

With the use of the new variables given by Eqs.~(\ref{transf1}) and (\ref
{transf2}), Eqs.~(\ref{e2}) and~(\ref{e1}) become
\begin{equation}
\frac{dq}{d\theta }=\mu -q,  \label{aut1}
\end{equation}
\begin{equation}
\frac{d\mu }{d\theta }=-\frac{\left( 2\mu +p\right) \left[ q+\frac{1}{3}%
\left( \mu +2p\right) \right] }{1-q}-2\frac{dp}{d\theta }+2\mu -2p.
\label{aut2}
\end{equation}

Eqs.~(\ref{e2}) and~(\ref{e1}), or, equivalently,~(\ref{aut1}) and~(\ref
{aut2}), are called the structure equations of the vacuum on the brane \cite
{Ha03}. In order to close the system of equations (\ref{aut1}) and~(\ref
{aut2}) an ``equation of state'' $p=p\left( \mu \right) $, relating the
reduced dark radiation and the dark pressure terms, is needed.

The structure equations of the vacuum on the brane can be solved exactly in
two cases, corresponding to some simple equations of state of the dark
pressure. In the first case we impose the equation of state $2\mu +p=0$.
From Eq. (\ref{aut2}) we immediately obtain $\mu =Qe^{-2\theta }$, while Eq.
(\ref{aut1}) gives $q\left( \theta \right) =-Qe^{-2\theta }+U_{0}e^{-\theta
} $, where $Q$ and $U_{0}=2GM$ are arbitrary constants of integration.
Therefore we obtain the vacuum brane solution
\begin{equation}
U=-\frac{P}{2}=\frac{Q}{3\alpha }\frac{1}{r^{4}},
\end{equation}
\begin{equation}
e^{-\lambda }=e^{\nu }=1-\frac{2GM}{r}+\frac{Q}{r^{2}}.
\end{equation}

This solution was first obtained in \cite{Da00}, and therefore corresponds
to an equation of state of the dark pressure of the form $P=-2U$. The second
case in which the vacuum structure equations can be integrated exactly
corresponds to the equation of state $\mu +2p=0$. Then Eq. (\ref{aut2})
gives $q=2/3$, and the corresponding solution of the gravitational field
equations on the brane is \cite{Ha03}
\begin{equation}
U=-2P=\frac{2}{9\alpha r^{2}}, e^{\nu }=C_{0}r^{2},e^{-\lambda }=\frac{1}{3}.
\end{equation}

This solution corresponds to an equation of state of the dark pressure of
the form $P=-U/2$.

\section{Kosambi-Cartan-Chern (KCC) theory and Jacobi stability}

\label{kcc}

We recall the basics of KCC-theory to be used in the sequel. Our exposition
follows \cite{Sa05}. 

Let $\mathcal{M}$ be a real, smooth $n$-dimensional manifold and let $T%
\mathcal{M}$ be its tangent bundle. Let $\left( x^{i}\right) =\left(
x^{1},x^{2},...,x^{n}\right) $, $\left( y^{i}\right) =\left(
y^{1},y^{2},...,y^{n}\right) $ and the time $t$ be a $2n+1$ coordinates
system of an open connected subset $\Omega $ of the Euclidian $(2n+1)$
dimensional space $R^{n}\times R^{n}\times R^{1}$, where
\begin{equation}
y^{i}=\left( \frac{dx^{1}}{dt},\frac{dx^{2}}{dt},...,\frac{dx^{n}}{dt}%
\right).
\end{equation}

We assume that $t$ is an absolute invariant, and therefore the only
admissible change of coordinates will be
\begin{equation}
\tilde{t}=t,\tilde{x}^{i}=\tilde{x}^{i}\left( x^{1},x^{2},...,x^{n}\right)
,i\in \left\{1 ,2,...,n\right\} .  \label{ct}
\end{equation}

The equations of motion of a dynamical system can be derived from a
Lagrangian $L$ via the Euler-Lagrange equations,
\begin{equation}
\frac{d}{dt}\frac{\partial L}{\partial y^{i}}-\frac{\partial L}{\partial
x^{i}}=F_{i},i=1,2,...,n,  \label{EL}
\end{equation}
where $F_{i}$, $i=1,2,...,n$, is the external force \cite{MHSS}. The triplet
$\left( M,L,F_{i}\right) $ is called a Finslerian mechanical system \cite
{MiFr05}. For a regular Lagrangian $L$, the Euler-Lagrange equations given
by Eq. (\ref{EL}) are equivalent to a system of second-order differential
equations
\begin{equation}
\frac{d^{2}x^{i}}{dt^{2}}+2G^{i}\left( x^{j},y^{j},t\right)
=0,i\in \left\{ 1,2,...,n\right\} ,  \label{EM}
\end{equation}
where each function $G^{i}\left( x^{j},y^{j},t\right) $ is $C^{\infty }$ in
a neighborhood of some initial conditions $\left( \left( x\right)
_{0},\left( y\right) _{0},t_{0}\right) $ in $\Omega $. The system given by
Eq.~(\ref{EM}) is equivalent to a vector field (semispray) $S$, where
\begin{equation}
S=y^{i}\frac{\partial }{\partial x^{i}}-2G^{i}\left( x^{j},y^{j},t\right)
\frac{\partial }{\partial y^{i}},
\end{equation}
which determines a non-linear connection $N_{j}^{i}$ defined as \cite{MHSS}
\begin{equation}
N_{j}^{i}=\frac{\partial G^{i}}{\partial y^{j}}.
\end{equation}


More generally, one can start from an arbitrary system of
second-order differential equations on the form (\ref{EM}), where
no \textit{a priori} given Lagrangean function is assumed, and
study the behavior of its trajectories by analogy with the
trajectories of the Euler-Lagrange equations.

For a non-singular coordinate transformations given by Eq. (\ref{ct}), we
define the KCC-covariant differential of a vector field $\xi ^{i}(x)$ on the
open subset $\Omega \subseteq R^{n}\times R^{n}\times R^{1}$ as \cite
{An93,An00,Sa05,Sa05a}
\begin{equation}
\frac{D\xi ^{i}}{dt}=\frac{d\xi ^{i}}{dt}+N_{j}^{i}\xi ^{j}.  \label{KCC}
\end{equation}

For $\xi ^{i}=y^{i}$ we obtain
\begin{equation}
\frac{Dy^{i}}{dt}=N_{j}^{j}y^{j}-2G^{i}=-\epsilon ^{i},
\end{equation}
where the contravariant vector field $\epsilon ^{i}$ on $\Omega $ is called
the first KCC invariant.

Let us now vary the trajectories $x^{i}(t)$ of the system (\ref{EM}) into
nearby ones according to
\begin{equation}
\tilde{x}^{i}\left( t\right) =x^{i}(t)+\eta \xi ^{i}(t),  \label{var}
\end{equation}
where $\left| \eta \right| $ is a small parameter and $\xi ^{i}(t)$ are the
components of some contravariant vector field defined along the path $%
x^{i}(t)$. Substituting Eqs. (\ref{var}) into Eqs. (\ref{EM}) and taking the
limit $\eta \rightarrow 0$ we obtain the variational equations \cite
{An93,An00,Sa05,Sa05a}
\begin{equation}
\frac{d^{2}\xi ^{i}}{dt^{2}}+2N_{j}^{i}\frac{d\xi ^{j}}{dt}+2\frac{\partial
G^{i}}{\partial x^{j}}\xi ^{j}=0.  \label{def}
\end{equation}

By using the KCC-covariant differential we can write Eq. (\ref{def}) in the
covariant form
\begin{equation}
\frac{D^{2}\xi ^{i}}{dt^{2}}=P_{j}^{i}\xi ^{j},  \label{JE}
\end{equation}
where we have denoted
\begin{equation}
P_{j}^{i}=-2\frac{\partial G^{i}}{\partial x^{j}}-2G^{l}G_{jl}^{i}+ y^{l}%
\frac{\partial N_{j}^{i}}{\partial x^{l}}+N_{l}^{i}N_{j}^{l}+\frac{\partial
N_{j}^{i}}{\partial t},
\end{equation}
and $G_{jl}^{i}\equiv \partial N_{j}^{i}/\partial y^{l}$ is called the
Berwald connection \cite{An93,MHSS,Sa05,Sa05a}. Eq. (\ref{JE}) is called the
Jacobi equation, and $P_{j}^{i}$ is called the second KCC-invariant or the
deviation curvature tensor. When the system (\ref{EM}) describes the
geodesic equations in either Riemann or Finsler geometry, Eq. (\ref{JE}) is
the Jacobi field equation.

The third, fourth and fifth invariants of the system (\ref{EM}) are given by
\cite{An00}
\begin{equation}
P_{jk}^{i}\equiv \frac{1}{3}\left( \frac{\partial P_{j}^{i}}{\partial y^{k}}-%
\frac{\partial P_{k}^{i}}{\partial y^{j}}\right) ,P_{jkl}^{i}\equiv \frac{%
\partial P_{jk}^{i}}{\partial y^{l}},D_{jkl}^{i}\equiv \frac{\partial
G_{jk}^{i}}{\partial y^{l}}.
\end{equation}

The third invariant is interpreted as a torsion tensor, while the fourth and
fifth invariants are the Riemann-Christoffel curvature tensor, and the
Douglas tensor, respectively \cite{An00}. In a Berwald space these tensors
always exist, and they describe the geometrical properties of a system of
second-order differential equations.

In many physical applications we are interested in the behavior of the
trajectories of the system (\ref{EM}) in a vicinity of a point $x^{i}\left(
t_{0}\right) $, where for simplicity one can take $t_{0}=0$. We will
consider the trajectories $x^{i}=x^{i}(t)$ as curves in the Euclidean space $%
\left( R^{n},\left\langle .,.\right\rangle \right) $, where $\left\langle
.,.\right\rangle $ is the canonical inner product of $R^{n}$. As for the
deviation vector $\xi $ we assume that it satisfies the initial conditions $%
\xi \left( 0\right) =O$ and $\dot{\xi}\left( 0\right) =W\neq O$, where $O\in
R^{n}$ is the null vector \cite{Sa05,Sa05a}.

For any two vectors $X,Y\in R^{n}$ we define an adapted inner product $%
\left\langle \left\langle .,.\right\rangle \right\rangle $ to the deviation
tensor $\xi $ by $\left\langle \left\langle X,Y\right\rangle \right\rangle
:=1/\left\langle W,W\right\rangle \cdot \left\langle X,Y\right\rangle $. We
also have $\left| \left| W\right| \right| ^{2}:=\left\langle \left\langle
W,W\right\rangle \right\rangle =1$.

Thus, the focusing tendency of the trajectories around $t_{0}=0$ can be
described as follows: if $\left| \left| \xi \left( t\right) \right| \right|
<t^{2}$, $t\approx 0^{+}$, the trajectories are bunching together, while if $%
\left| \left| \xi \left( t\right) \right| \right| >t^{2}$, $t\approx 0^{+}$,
the trajectories are dispersing \cite{Sa05,Sa05a}. In terms of the deviation
curvature tensor the focusing tendency of the trajectories can be described
as follows: The trajectories of the system of equations (\ref{EM}) are
bunching together for $t\approx 0^{+}$ if and only if the real part of the
eigenvalues of $P_{j}^{i}\left( 0\right) $ are strictly negative, and they
are dispersing if and only if the real part of the eigenvalues of $%
P_{j}^{i}\left( 0\right) $ are strict positive \cite{Sa05,Sa05a}.

Based on these considerations we can define the Jacobi stability for a
dynamical system as follows \cite{An00,Sa05,Sa05a}:

\textbf{Definition:} If the system of differential equations (\ref{EM})
satisfies the initial conditions $\left| \left| x^{i}\left( t_{0}\right) -%
\tilde{x}^{i}\left( t_{0}\right) \right| \right| =0$, $\left| \left| \dot{x}%
^{i}\left( t_{0}\right) -\tilde{x}^{i}\left( t_{0}\right) \right| \right|
\neq 0$, with respect to the norm $\left| \left| .\right| \right| $ induced
by a positive definite inner product, then the trajectories of (\ref{EM})
are Jacobi stable if and only if the real parts of the eigenvalues of the
deviation tensor $P_{j}^{i}$ are strictly negative everywhere, and Jacobi
unstable, otherwise.

\begin{figure}[!ht]
\centering
\includegraphics[height=6cm,width=12cm]{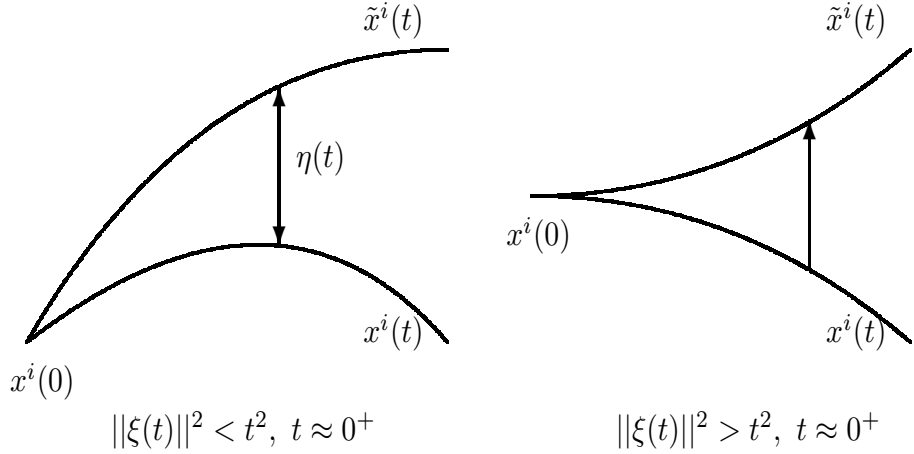} 
\caption{Behavior of trajectories near zero.}
\label{pict1}
\end{figure}

The focussing behavior of the trajectories near the origin is represented in
Fig.~\ref{pict1}.

\section{Linear stability analysis of the vacuum structure equations on the
brane for the linear equation of state of the dark pressure}

\label{linstab}

Since generally the structure equations of the vacuum on the brane cannot be
solved exactly, in this Section we shall analyze them by using methods from
the qualitative analysis of dynamical systems~\cite{BoHa07}, by closely
following the approach of ~\cite{boyce}. We consider the case in which the
dark pressure is proportional to the dark radiation, $P=\gamma U$, where $%
\gamma $ is an arbitrary constant, which can take both positive and negative
values. As we have seen in Section~\ref{Sect3}, several classes of exact
solutions of the vacuum gravitational field equations on the brane can be
described by an equation of state of this form. In the reduced variables $%
\mu $ and $p$ the linear equation of state is $p=\gamma \mu $, and the
structure equations of the gravitational field on the brane have the form
\begin{equation}
\frac{dq}{d\theta }=\mu -q,  \label{aut3a}
\end{equation}
\begin{equation}
\left( 1+2\gamma \right) \frac{d\mu }{d\theta }=2\left( 1-\gamma \right) \mu
-\frac{\left( \gamma +2\right) \mu \left[ q+\frac{1+2\gamma }{3}\mu \right]
}{1-q}.  \label{aut4a}
\end{equation}

Let us firstly analyze the special case where $\gamma =-1/2$. Then, by
virtue of the second Eq.~(\ref{aut4a}), we obtain two possible exact
solutions, either $q=\mu =0$ or $q=\mu =2/3$. The first of these solutions
correspond to the vanishing of all physical quantities, and therefore we can
discard it as unphysical. The second case corresponds to an exact solution,
which has been discussed in Section~\ref{Sect3}.

Let us henceforth assume that $\gamma \neq -1/2$, and rewrite the system of
equation into the following form,
\begin{equation}
\frac{dq}{d\theta }=-q+\mu ,  \label{aut3b}
\end{equation}
\begin{equation}
\frac{d\mu }{d\theta }=\frac{2(1-\gamma )}{1+2\gamma }\mu -\frac{\gamma +2}{%
1+2\gamma }\frac{\mu \left[ q+\frac{1+2\gamma }{3}\mu \right] }{1-q},
\label{aut4b}
\end{equation}
which is finally written as
\begin{equation}
\frac{d\xi }{d\theta }=A\xi +B,  \label{autuse}
\end{equation}
where we have denoted
\begin{equation}
\xi =
\begin{pmatrix}
q \\
\mu
\end{pmatrix}
,\quad A=
\begin{pmatrix}
-1 & 1 \\
0 & 2(1-\gamma )/(1+2\gamma )
\end{pmatrix}
,\quad B=
\begin{pmatrix}
0 \\
\frac{\gamma +2}{1+2\gamma }\frac{\mu \left[ q+\frac{1+2\gamma }{3}\mu
\right] }{1-q}
\end{pmatrix}
\end{equation}

The system of equations~(\ref{autuse}) has two critical points, $X_{0}=(0,0)$%
, and

\begin{equation}
X_{\gamma }=\left( \frac{3(1-\gamma )}{\gamma ^{2}+\gamma +7},\frac{%
3(1-\gamma )}{\gamma ^{2}+\gamma +7}\right) .  \label{critp}
\end{equation}

For $\gamma =1$, the two critical points of the system coincide. Depending
on the values of $\gamma $, these points lie in different regions of the
phase space plane~$(q,\mu )$.

Since the term $||B||/||\xi ||\rightarrow 0$ as $||\xi ||\rightarrow 0$, the
system of equations~(\ref{autuse}) can be linearized at the critical point $%
X_{0}$. The two eigenvalues of the matrix $A$ are given by $r_{1}=-1$ and $%
r_{2}=2(1-\gamma )/(1+2\gamma )$, and determine the characteristics of the
critical point $X_{0}$. For $\gamma \in (-\infty ,-1/2)\cup \lbrack 1,\infty
)$ both eigenvalues are negative and unequal. Therefore, for such values of $%
\gamma $ the point $X_{0}$ is an improper asymptotically stable node.

If $\gamma \in (-1/2,1)$, we find one positive and one negative eigenvalue,
which corresponds to an unstable saddle point at the point $X_0$. Moreover,
since the matrix $dA/d\xi(X_0)$ has real non-vanishing eigenvalues, the
point $X_0$ is hyperbolic. This implies that the properties of the
linearized system are also valid for the full non-linear system near the
point $X_0$. It should be mentioned however, that this first critical point
is the less interesting one from a physical point of view, since it
corresponds to the `trivial' case where both physical variables vanish.

As we have seen in Section~\ref{Sect3}, the structure equations can also be
solved exactly for the value $\gamma =-2$. In that case, the non-linear term
$B$ in Eq.~(\ref{autuse}) identically vanishes, and the system of equations
becomes a simple linear system of differential equations. For $\gamma =-2$
the two eigenvalues of $A$ are given by $r_{1}=-1$ and $r_{2}=-2$, and the
two corresponding eigenvectors are linearly independent. The general
solution can be written as follows
\begin{equation}
\xi _{\gamma =-2}=(q_{0}+\mu _{0})
\begin{pmatrix}
1 \\
0
\end{pmatrix}
e^{-\theta }+\mu _{0}
\begin{pmatrix}
-1 \\
1
\end{pmatrix}
e^{-2\theta },
\end{equation}
where $q_{0}=q(0)$ and $\mu _{0}=\mu (0)$. One can easily transform this
solution back into the radial coordinate $r$ form by using $\theta =\ln (r)$%
, thus obtaining
\begin{equation}
\mu _{\gamma =-2}=\frac{\mu _{0}}{r^{2}},\quad q_{\gamma =-2}=\frac{q_{0}}{r}%
+\mu _{0}\left( \frac{1}{r}-\frac{1}{r^{2}}\right) .
\end{equation}

Let us now analyze the qualitative behavior of the second critical point $%
X_{\gamma }$. To do this, one has to Taylor expand the right-hand sides of
Eqs.~(\ref{aut3b}) and~(\ref{aut4b}) around $X_{\gamma }$ and obtain the
matrix $\tilde{A}$ which corresponds to the system, linearized around $%
X_{\gamma }$. This linearization is again allowed since the resulting
non-linear term, $\tilde{n}$ say, also satisfies the condition $||\tilde{n}%
||/||\xi ||\rightarrow 0$ as $||\xi ||\rightarrow X_{\gamma }$. The
resulting matrix reads
\begin{equation}
\tilde{A}=
\begin{pmatrix}
-1 & 1 \\
\frac{3(-\gamma ^{2}+5\gamma -4)}{(2+\gamma )^{2}(1+2\gamma )} & \frac{%
\gamma -1}{2+\gamma }
\end{pmatrix}
,  \label{ew}
\end{equation}
and its two eigenvalues are given by
\begin{equation}
r_{\pm }=\frac{-3-6\gamma \pm \sqrt{16\gamma ^{4}+8\gamma ^{3}+132\gamma
^{2}-28\gamma -47}}{4\gamma ^{2}+10\gamma +4}.
\end{equation}

For $-0.5 < \gamma < 0.674865$ the argument of the square root becomes
negative and the eigenvalues complex. Moreover, the values $\gamma=-1/2$ and
$\gamma=-2$ have to be excluded, since the eigenvalues in Eq.~(\ref{ew}) are
not defined in these cases. However, both cases have been treated separately
above.

If $\gamma \in (-\infty ,-1/2)\cup (1,\infty )$, then $X_{\gamma }$
corresponds to an unstable saddle point and for $\gamma \in (0.674865,1)$ it
corresponds to an asymptotically stable improper node. More interesting is
the parameter range $\gamma \in (-1/2,0.674865)$, where the eigenvalues
become complex, however, their real parts are negative definite. Hence, for
those values we find an asymptotically stable spiral point at $X_{\gamma }$.
Since this point is also a hyperbolic point, the described properties are
also valid for the non-linear system near that point.

\begin{table}[tbp]
\begin{center}
\begin{tabular}{|c|ccccccccc|}
\hline
$\gamma$ & {$-\infty $} &  & -0.5 &  & 0.67 &  & 1 &  & +$\infty$ \\ \hline
{$r_{\pm}$} &  & real & \big| & complex & \big| & real & \big| & real &  \\
\hline
&  &  & \big| & Re {$r_{\pm}<0$} & \big| &  & \big| &  &  \\ \hline
{$r_{+}$} &  & + & \big| &  & \big| & -- & \big| & -- &  \\ \hline
{$r_{-}$} &  & -- & \big| &  & \big| & -- & \big| & + &  \\ \hline
{$X_{\gamma}$} &  & Saddle & \big| & Stable spiral & \big| & Stable node & %
\big| & Saddle &  \\ \hline
\end{tabular}
\end{center}
\caption{Linear stability of the stable point $X_{\protect\gamma}$.}
\label{table1}
\end{table}

The behavior of the trajectories is shown, for $\gamma =-1$ and $\gamma =0.4$%
, in Fig.~\ref{f2}. The figures show the attracting or repelling
character of the steady states, respectively. The results of the
linear stability analysis of the critical points $X_{\gamma}$ are
summarized in Table~\ref{table1}.
\begin{figure}[!ht]
\centering
\includegraphics[height=8cm,width=7.5cm,angle=270]{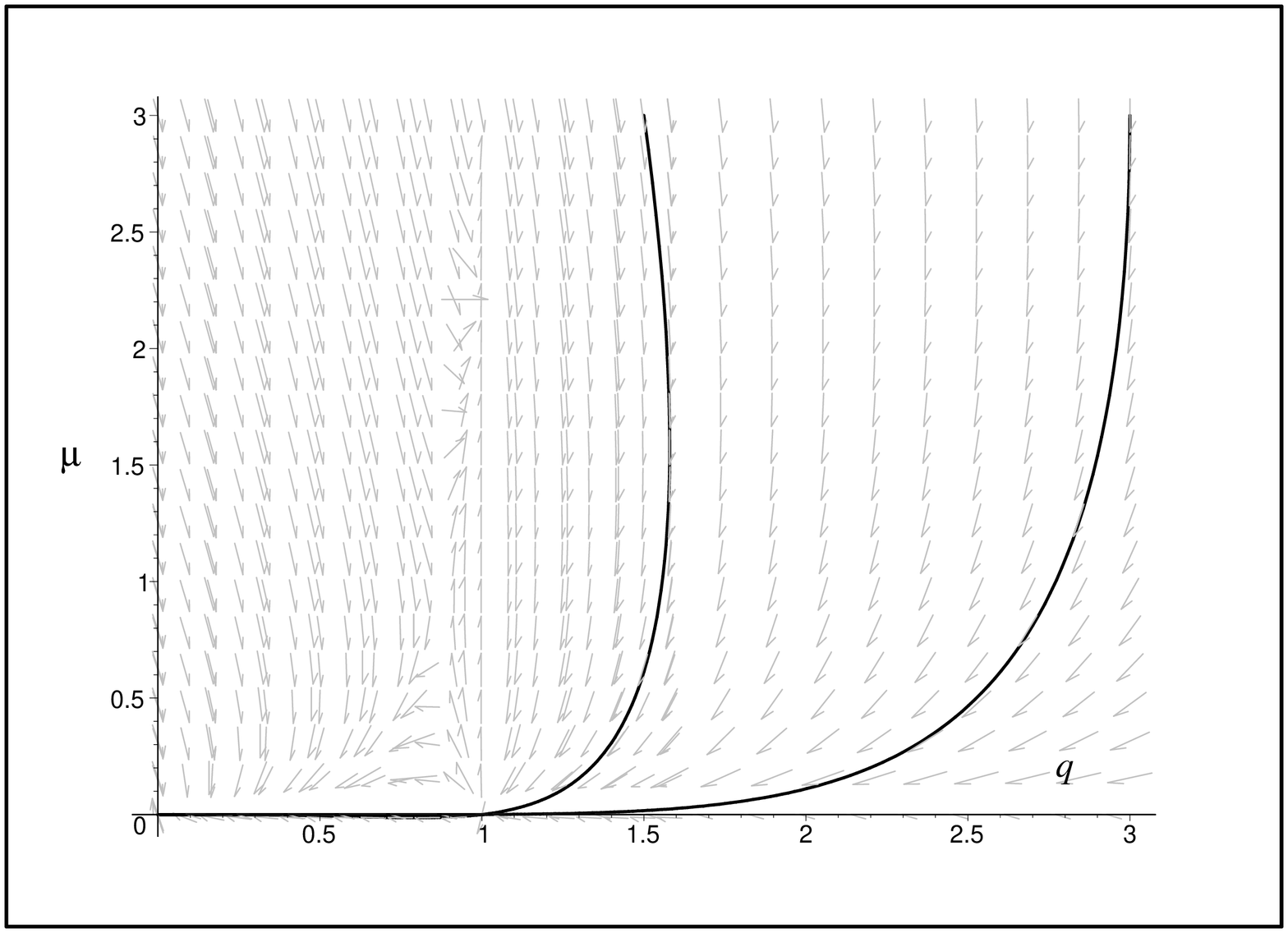} %
\includegraphics[height=8cm,width=7.5cm,angle=270]{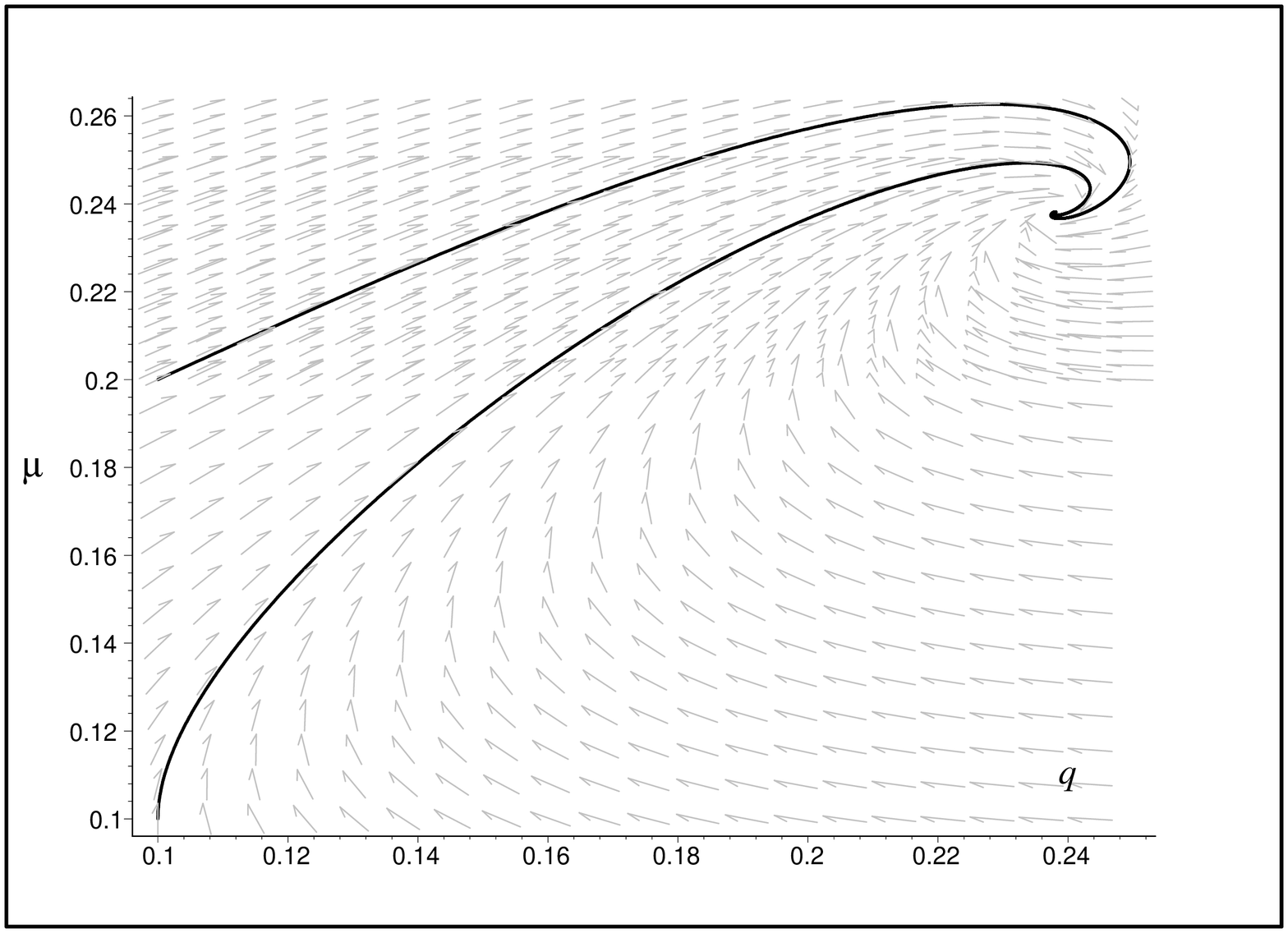}
\caption{Behavior of the trajectories of the structure equations of the
vacuum on the brane near the critical points for $\protect\gamma =-1$ (left
figure) and $\protect\gamma =0.4$ (right figure).}
\label{f2}
\end{figure}

\section{Jacobi stability analysis of the vacuum gravitational field
equations in the brane world models}

Since from Eq. (\ref{aut3a}) we can express $\mu $ as $\mu =q+dq/d\theta $,
upon substitution in Eq. (\ref{aut4a}) we obtain for $q$ the following
second order differential equation
\begin{equation}
\begin{split}
\frac{d^{2}q}{d\theta ^{2}} & + \frac{1}{3\left( 1+2\gamma \right) \left(
1-q\right)} \Bigl[6\left( \gamma -1\right) q+2\left( \gamma ^{2}+\gamma
+7\right) q^{2}+3\left( 4\gamma -1\right) \frac{dq}{d\theta } \\
& +\left( 4\gamma ^{2}+\gamma +13\right) q\frac{dq}{d\theta }+\left( 2\gamma
^{2}+5\gamma +2\right) \left( \frac{dq}{d\theta }\right) ^{2} \Bigr]=0,
\end{split}
\label{jac1}
\end{equation}
which can now be studied by means of KCC theory.

By denoting $x=q$ and $dq/d\theta =dx/d\theta =y$, Eq. (\ref{jac1}) can be
written as
\begin{equation}
\frac{d^{2}x}{d\theta ^{2}}+2G^{1}\left( x,y\right) =0,
\end{equation}
where
\begin{equation}
\begin{split}
G^{1}\left( x,y\right) & =\frac{1}{6\left( 1+2\gamma \right) \left(
1-x\right)} \Bigl[6\left( \gamma -1\right) x+2\left( \gamma ^{2}+\gamma
+7\right) x^{2}+3\left( 4\gamma -1\right) y \\
& +\left( 4\gamma ^{2}+\gamma +13\right) xy+\left( 2\gamma ^{2}+5\gamma
+2\right) y^{2} \Bigr]
\end{split}
\end{equation}

As a first step in the KCC stability analysis of the vacuum field equations
on the brane we obtain the nonlinear connection $N_{1}^{1}$ associated to
Eq. (\ref{jac1}), and which is given by
\begin{equation}
N_{1}^{1}=\frac{\partial G^{1}}{\partial y}=\frac{3\left( 4\gamma -1\right)
+\left( 4\gamma ^{2}+\gamma +13\right) x+2\left( 2\gamma ^{2}+5\gamma
+2\right) y}{6\left( 1+2\gamma \right) \left( 1-x\right) }.
\end{equation}

The Berwald connection can be obtained as
\begin{equation}
G_{11}^{1}=\frac{\partial N_{1}^{1}}{\partial y}=\frac{ 2\gamma ^{2}+5\gamma
+2 }{3\left( 1+2\gamma \right) \left( 1-x\right) }.
\end{equation}

Finally, the second KCC invariant or the deviation curvature tensor $%
P_{1}^{1}$, defined as
\begin{equation}
P_{1}^{1}=-2\frac{\partial G^{1}}{\partial x}-2G^{1}G_{11}^{1}+y\frac{%
\partial N_{1}^{1}}{\partial x}+N_{1}^{1}N_{1}^{1},
\end{equation}
reads now 
\begin{equation}
P_{1}^{1}\left( x,y\right) =\frac{27-2\left[ 61+57\gamma +4\gamma ^{2}\left(
9+2\gamma \right) \right] x+3\left( 5+\gamma \right) ^{2}x^{2}-2\left(
2+\gamma \right) \left( 1+2\gamma \right) \left( 5+4\gamma \right) y}{%
12\left( 1-x\right) ^{2}\left( 1+2\gamma \right) ^{2}}.
\end{equation}

Taking into account that $x=q$ and $y=\mu -q$, we obtain $P_{1}^{1}$ in the
initial variables as 
\begin{equation}
P_{1}^{1}\left( q,\mu \right) =\frac{27-6\left( 17+8\gamma +2\gamma
^{2}\right) q+3\left( 5+\gamma \right) ^{2}q^{2}-2\left( 2+\gamma \right)
\left( 1+2\gamma \right) \left( 5+4\gamma \right) \mu }{12\left( 1-q\right)
^{2}\left( 1+2\gamma \right) ^{2}}.
\end{equation}

Evaluating $P_{1}^{1}\left( q,\mu \right) $ at the critical point $X_{\gamma
}$, given by Eq. (\ref{critp}), ew obtain 
\begin{equation}
P_{1}^{1}\left( X_{\gamma }\right) = \frac{8\gamma^3 + 66\gamma -47} {%
4\left(2+\gamma \right)^2\left(1+2\gamma \right)}.
\end{equation}

The plot of the function $P_{1}^{1}\left( X_{\gamma }\right) $ is
represented in Fig. \ref{f3}. 
\begin{figure}[!ht]
\centering
\includegraphics[height=8cm,width=12cm]{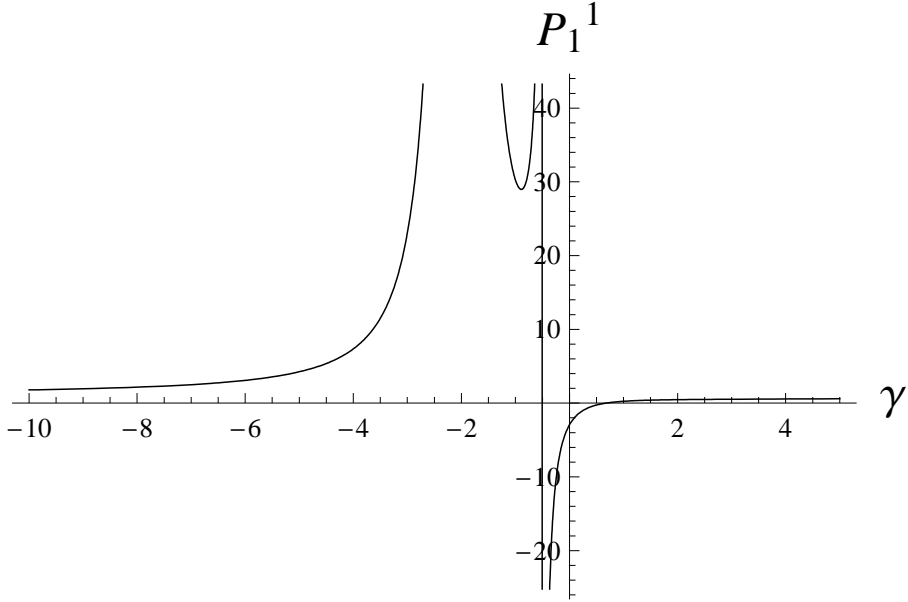}
\caption{The deviation curvature tensor $P_{1}^{1}\left( X_{\protect\gamma
}\right) $ as a function of $\protect\gamma $.}
\label{f3}
\end{figure}

Using the discussion of Section~\ref{linstab}, our main results on the
Linear stability and Jacobi stability of the critical point $X_{\gamma}$ of
the vacuum field equations in the brane world models can be summarized in
Table~\ref{table2}.

\begin{table}[tbp]
\begin{center}
\begin{tabular}{|c|ccccccccc|}
\hline
$\gamma$ & {$-\infty $} &  & -0.5 &  & 0.67 &  & 1 &  & +$\infty$ \\ \hline
{$P_1^1(X_{\gamma})$} &  & + & \big| & -- & \big| & + & \big| & + &  \\
\hline
{Linear stability } &  & Saddle & \big| & Stable & \big| & Stable & \big| &
Saddle &  \\
{of $X_{\gamma}$} &  & point & \big| & spiral & \big| & node & \big| & point
&  \\ \hline
{Jacobi stability } &  & Jacobi & \big| & Jacobi & \big| & Jacobi & \big| &
Jacobi &  \\
{of $X_{\gamma}$} &  & unstable & \big| & stable & \big| & unstable & \big|
& unstable &  \\ \hline
\end{tabular}
\end{center}
\caption{Linear and Jacobi stability of the stable point $X_{\protect\gamma}
$.}
\label{table2}
\end{table}

\section{Discussions and final remarks}

In the present paper we have considered the stability properties of the
vacuum gravitational field equations in the brane world models. For the
analysis of the stability we have used two methods, the Lyapunov (linear)
stability analysis, and the so-called Jacobi stability analysis, or the KCC
theory. The study of the stability has been done by analyzing the behavior
of the steady state $X_{\gamma}$ of the structure equations of the vacuum on
the brane. The Lyapunov stability analysis involves the linearization of the
dynamical system via the Jacobian matrix of a non-linear system, while the
KCC theory addresses the Lyapunov stability of a whole trajectory in a
tubular region \cite{Sa05}.

By using the KCC theory we have shown that the vacuum on the brane is Jacobi
unstable for most of the values of the parameter $\gamma $. The stability
region is reduced to a very narrow range of $\gamma $, $\gamma
\in\left(-0.5,0.67\right)$. For all other values of $\gamma $ the vacuum on
the brane is unstable, in the sense that the trajectories of the structure
equations will disperse when approaching the origin of the coordinate system.

In a previous paper (\cite{Sa05}) we have regarded the Jacobi stability of a
dynamical system as the \textit{robustness} of the system to small
perturbations of the whole trajectory. This is a very convenable way of
regarding the resistance of limit cycles to small perturbation of
trajectories. On the other hand, we may regard the Jacobi stability for
other types of dynamical systems (like the one in the present paper) as the
resistance of a whole trajectory to the onset of chaos due to small
perturbations of the whole trajectory. This interpretation is based on the
generally accepted definition of chaos, namely a compact manifold $M$ on
which the geodesic trajectories deviate exponentially fast. This is
obviously related to the curvature of the base manifold (see section IV).
The Jacobi (in)stability is a natural generalization of the (in)stability of
the geodesic flow on a differentiable manifold endowed with a metric
(Riemannian or Finslerian) to the non-metric setting. In other words, we may
say that Jacobi unstable trajectories of a dynamical system behave
chaotically in the sense that after a finite interval of time it would be
impossible to distinguish the trajectories that were very near each other at
an initial moment.

We have found that there is a good correlation between the linear stability
of the critical point $X_{\gamma}$ and the robustness of the corresponding
trajectory to small perturbations. Indeed, for small values of the parameter
$\gamma$ the saddle point is also Jacobi unstable ($\gamma<-0.5$ and $%
\gamma>1$), while the stable spiral obtained for
$-0.5<\gamma<0.674865$ is also robust to a small perturbation of
all trajectories.

It is interesting to remark that for the interval $0.674865<\gamma<1$, the
stable node is actually Jacobi unstable. In other words, even the system
trajectories are attracted by the critical point $X_{\gamma}$ one has to be
aware of the fact that they are not stable to small perturbation of the
whole trajectory. This means that one might witness chaotic behavior of the
system trajectories before they enters in a neighborhood of $X_{\gamma}$. We
have here a sort of stability artifact that cannot be found without using
the powerful method of Jacobi stability analysis.

\acknowledgments

The work of T.~H.~was supported by an RGC grant of the government of the
Hong Kong SAR.

\end{document}